# MEASUREMENT OF THE ION DRAG FORCE IN A COMPLEX DC-PLASMA USING THE PK-4 EXPERIMENT

M.H.THOMA[1], H. HÖFNER[1], S. A. KHRAPAK[1], M. KRETSCHMER[1],
R.A. QUINN[1], S. RATYNSKAIA[1], G.E. MORFILL[1], A. USACHEV[2],
A. ZOBNIN[2], O. PETROV[2], V. FORTOV[2]

[1]Centre for Interdisciplinary Plasma Science, Max-Planck-Institut
für extraterrestrische Physik, P.O. Box 1312, D-85741 Garching, Germany,
thoma@mpe.mpg.de
[2]Institute for High Energy Densities, Russian Academy of Sciences,
Izhorskaya, 13/19, 127412, Moscow, Russia

**Abstract:** The force on a microparticle in a complex plasma by streaming ions, the so-called ion drag force, is not well known. However, it is important for the understanding of interesting phenomena in complex plasmas such as the void formation under microgravity conditions. The PK-4 experiment, which is developed for a later use on board of the International Space Station, is ideally suited for investigating this problem. In this experiment a complex DC-plasma is created in a glass tube in which the microparticles flow from the cathode to the anode. Measuring the microparticle velocities, the forces on the particles for different particle sizes, pressures, and DC-currents can be extracted by assuming force balance. Experiments have been performed in the laboratory as well as under microgravity using parabolic flights. The results of these experiments will be presented and compared to theoretical predictions.

Complex or dusty plasmas are low-temperature plasmas containing highly charged micron size particles. Mostly rf plasma chambers have been used to study the properties of complex plasmas, in particular regular structures in the microparticle component, the so-called plasma crystal [1]. Beside laboratory experiments, investigations under microgravity conditions are performed to avoid the disturbing effects of gravity on the microparticles [2]. Also a few experiments in a dc plasma chamber were conducted in the laboratory as well as under zero gravity [3]. For this purpose elongated glass cylinders with dc electrodes at their ends have been used. In addition external rf coils or electrodes can be applied to study the system in a combined dc/rf plasma.

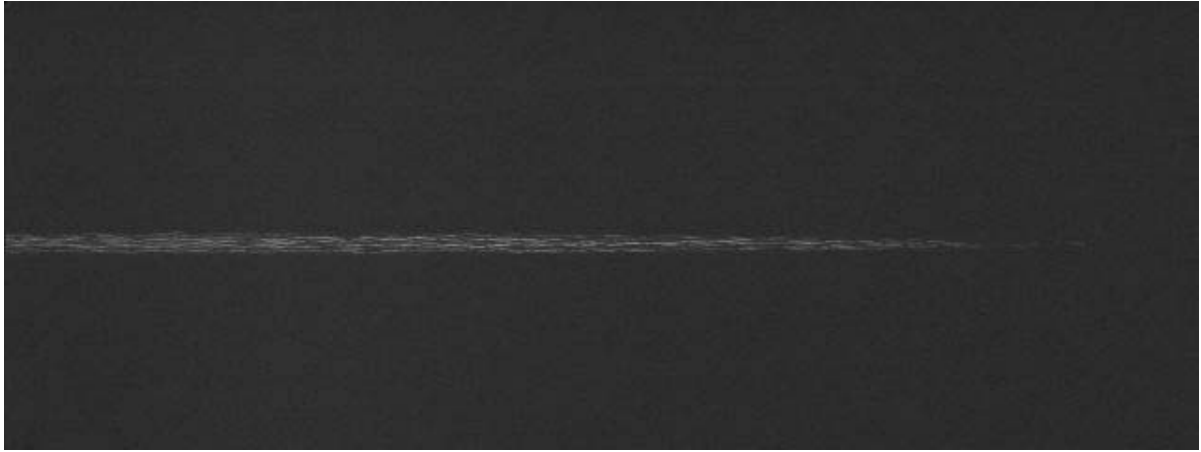

Fig. 1: Photographic image of the microparticle flow in the PK-4 plasma chamber.

PK-4 (``Plasmakristall-4'') is an experimental project for investigating complex dc and combined dc/rf plasmas under microgravity conditions on board of the International Space Station ISS after 2007 [4]. At the moment a laboratory experiment set-up is developed and first experiments are performed in the laboratory and in parabolic flight campaigns in a German-Russian collaboration. The PK-4 set-up is especially suited to study the streaming of microparticles in the liquid phase of the complex plasma along the glass tube (Fig. 1). Measurements of the particle velocities for different pressures, dc currents, and particle sizes together with probe measurements of the plasma parameters allow the determination of the various forces acting on the particles. The electric charge of the particles has also been extracted in this way, showing that the standard OML theory [5] overestimates the charge up to a factor of 5 [6]. Also dust wave instabilities have been observed below a pressure threshold allowing an independent determination of the particle charge [6]. Furthermore, cloud collisions, soliton-like waves, and Lavalle nozzle simulations have been studied.

The aim of the present work is to determine the ion drag force acting on the microparticles by the interaction between the particle and the ion components streaming against each other. This force is the least understood force acting on the particles. Its knowledge is of great importance for interesting phenomena in complex plasmas such as void formation under microgravity conditions [7]. At the moment there is a controversy about the ion drag force: the model by Barnes et al. [8] assumes that "... no ion interaction with the particles occurs outside of a Debye length" according to the standard Coulomb scattering theory. However, in complex plasmas the range of the ion-microparticle interaction is usually larger than the Debye screening length. Hence standard Coluomb scattering theory is not applicable.

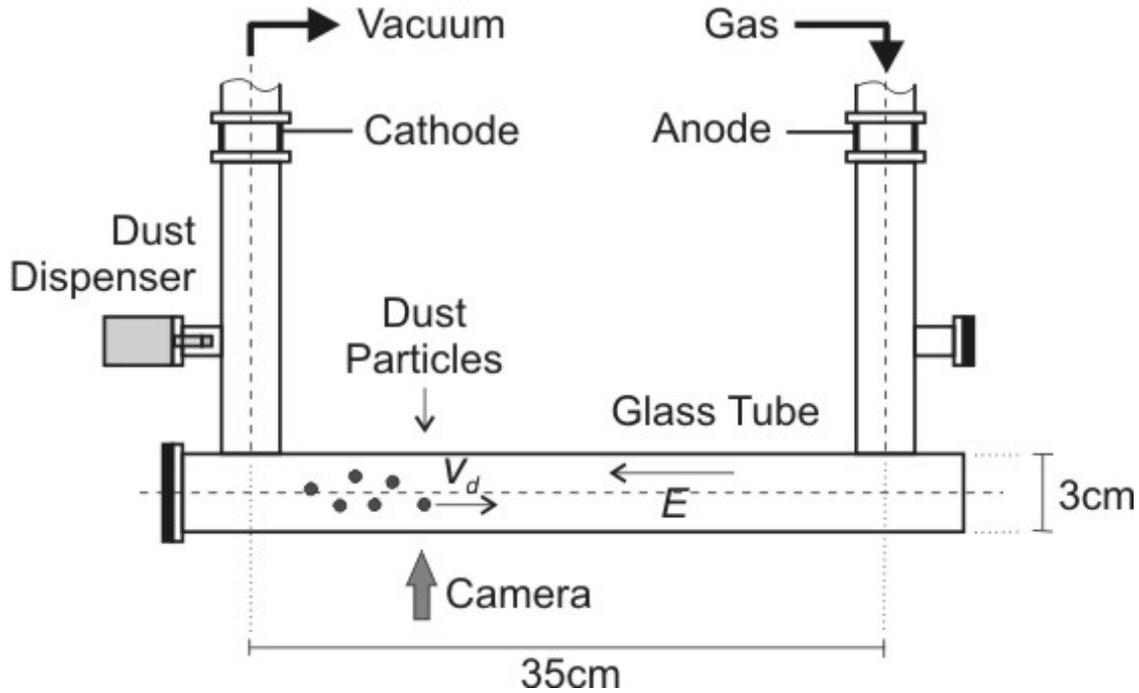

Fig.2: Sketch of the PK-4 chamber.

As shown by Khrapak et al. [9] this fact can lead to a strong enhancement of the ion drag force compared to the model by Barnes. Investigating the trajectories of particles falling through a rf discharge plasma, Zafiu et al. [10,11] concluded that the ion drag force is given by Barnes' formula if the *electron* Debye length is used there. However, in a bulk plasma with a subthermal ion drift velocity the Debye length is usually given by the *ion* Debye length leading to a significantly smaller ion drag force using Barnes' formula [12].

In the PK-4 plasma chambers (see Fig. 2) filled with neon at the Max-Planck-Institute for Extraterrestrial Physics (MPE) and the Institute for High Energy Densities (IHED) the velocities of particles with different sizes (diameter from 1.28 to 11 µm) have been measured by following the tracks of the particles, illuminated by a laser, on a film. The ion drag force $F_i$ follows from the force balance equation

$$F_i = F_E - F_n + F_g \tag{1}$$

(or $F_i = F_E + F_n - F_g$ if the vertical tube configuration is used, the cathode is on top, and $F_g > F_E$.) This equation assumes that the particles have reached their final velocity due to friction on the neutral gas in the field of view of the camera. This is true since the final velocity of the particles is achieved already within a fraction of a second (even for the largest particles and smallest pressures used) whereas the particles enter the field of view a few seconds after their injection.

The first force on the right hand side, $F_E = Q E$, is the electric force on a particle with charge $Q$ caused by the longitudinal electric dc field $E$ (about 2 V/cm), which has been measured at IHED using a Langmuir probe. The particle charge $Q$ was taken from a molecular dynamics calculations by Zobnin et al. [14], which give results in good agreement with PK-4 measurements [6]. To avoid a reduction of the charge by high particle densities (Havnes effect) we worked with the lowest dispenser settings to obtain particle densities as small as possible.

The second force, $F_n = \gamma v_p$, is the neutral drag friction with the Epstein coefficient $\gamma$ and the measured particle velocity $v_p$. The Epstein coefficient is proportional to a coefficient $\delta$ ranging from 1 for elastic scattering to 1.4 for diffuse reflection with full accommodation of the neutral gas atoms on the particles. However, falling particle experiments in the PK-4 chamber in neon gas without plasma indicated that the Epstein coefficient should be larger ($\delta = 1.6$), which was also found by Liu et al. [13]. This can be explained possibly by a smaller size (reduction of the diameter by about 10%) of the melamine formaldehyde particles than given by the manufacturer. Indeed a microscopic inspection of the particles by Liu et al. [13] and at IHED showed such a reduction, which should be considered also for other forces and for other high precision experiments.

The gravitational force $F_g$ is negligible in the parabolic flight experiments, where particles with nominal diameter of 6.86 μm were used, and in the laboratory if the tube is in the horizontal configuration. However, in the latter case only small particles with diameter from 1.28 to 3.42 μm can be injected into the tube in a way that their flow through the tube can be observed. For larger particles the weak ambipolar radial electric field is not sufficient to compensate gravity and the particles are lost to the tube walls. Therefore it is necessary to use a vertical configuration, where gravity is opposite to the electric force, i.e., the cathode is on top. The negatively charged particles stream downwards as gravity is always dominating for large particle sizes. However, due to the complete domination of the gravitational force it is difficult to determine weak forces such as the ion drag force, showing the importance of microgravity experiments. Here we will concentrate only on laboratory experiments with smaller particles in the horizontal tube configuration and parabolic flight experiments for large particles.

The microparticles were illuminated by a laser and recorded on a film with 120 frames per second by a CCD camera. In Table 1 the velocities of the microparticles, measured from their track lengths on single images of the film (exposure time 8ms) or from the particle positions in subsequent images, are shown. About 20 tracks for each parameter set were examined, leading to the statistical errors given in Table 1. In the case of the largest particles (diameter 6.86 µm) used in parabolic flights a larger velocity variation was observed. All these measurements were performed at a current of 1 mA.

| p[Pa] | 1.28 | 1.95 | 2.55 | 3.42 | 6.86 |
|---|---|---|---|---|---|
| 20 | 12.8 +/- 0.20 | 7.98 +/- 0.12 | 8.01 +/- 0.06 | 5.96 +/- 0.06 | 1.95 +/- 0.65 |
| 40 | 6.65 +/- 0.16 | 3.84 +/- 0.04 | 3.62 +/- 0.04 | 3.40 +/- 0.03 | 1.50 +/- 0.30 |
| 60 | 3.45 +/- 0.08 | 2.36 +/- 0.02 | 2.15 +/- 0.02 | 2.18 +/- 0.02 | 0.77 +/- 0.05 |

Table 1: Measured particle velocities in cm/s at 3 different pressures and various particle diameters from 1.28 to 6.86 µm at a current of 1 mA.

In Table 2 the charges of the microparticles taken from the molecular dynamics simulations of Ref. [14] are listed. Here we assumed the particle sizes as given by the manufacturer and plasma parameters (electron temperature and density) as measured by using the Langmuir probe at IHED.

| p[Pa] | 1.28 | 1.95 | 2.55 | 3.42 | 6.86 |
|---|---|---|---|---|---|
| 20 | 3000 | 4600 | 6000 | 8200 | 16900 |
| 40 | 2400 | 3800 | 5100 | 7000 | 14900 |
| 60 | 2100 | 3300 | 4300 | 5900 | 12300 |

Table 2: Microparticle charges from molecular dynamics simulations.

In Table 3 the ion drag force, derived from (1) with $F_g = 0$, is shown. We assumed the particle size as given by the manufacturer and used an accommodation coefficient $\delta = 1.6$ as found in the falling particle experiments. We do not give any error in the ion drag force because the main uncertainty comes from uncertainties in the theoretical predicted charge and the measured electric field which are difficult to estimate.

| p[Pa] | 1.28 | 1.95 | 2.55 | 3.42 | 6.86 |
|---|---|---|---|---|---|
| 20 | 6.4 | 10.1 | 11.3 | 15.6 | 25 |
| 40 | 4.4 | 7.7 | 9.1 | 10.3 | 21 |
| 60 | 4.1 | 6.5 | 7.4 | 7.3 | 16 |

Table 3: Experimentally measured ion drag force in $10^{-14}$ N.

In the tables below the measured ion drag force is compared to theoretical predictions. Here $F_i(1)$ indicates the prediction by Khrapak et al. [9], where the ion drag force is given by

$$F_i = \frac{2\sqrt{2\pi}}{3} n_i m_i v_{Ti} u \rho_0^2 \Lambda. \tag{2}$$

Here the ion density $n_i$ is assumed to be equal to the electron density measured by the Langmuir probe ($n_i = 1.6 - 2.6 \times 10^8$ cm$^{-3}$), the ion mass (neon) is given by $m_i = 3.35 \times 10^{-26}$ kg and the thermal ion velocity by $v_{Ti}=(kT_i/m_i)^{1/2} = 351$ m/s assuming for ion temperature $T_i = 300$ K. The ion drift velocity according to Raizer [15] is approximately $u$[m/s] $= 3900\ E$[V/cm]/$p$[Pa], the Coulomb radius is defined by $\rho_0 = Ze^2/kT_i$ with the particle charge $Z$, and $\Lambda$ is the modified Coulomb logarithm

$$\Lambda = 2 \int_0^\infty dx\, e^{-x} \ln \frac{2\lambda_D x + \rho_0}{2ax + \rho_0} \tag{3}$$

with particle radius $a$ and ion Debye length $\lambda_D$ [µm] $= 120/(n_i[10^8$ cm$^{-3}])^{1/2}$. The ion Debeye length ranges from 74 µm at 60 Pa to 95 µm at 20 Pa. In (2) only the orbital part of the ion drag force due to Coulomb scattering is considered as the collection part is negligible for our choice of parameters. The ratio of the Coulomb radius to the ion Debye length in our experiment is between 1.4 for the 1.28 µm particles up to 10.2 for the 6.86 µm particles, indicating that the standard Coulomb scattering theory is not valid [9].

$F_i(2)$ and $F_i(3)$ follow from Barnes' formula [8],

$$F_i = \pi n_i m_i u v_s (b_c^2 + 4b_{\pi/2}^2 \Gamma), \tag{4}$$

where $v_s = [8kT_i/(\pi m_i) + u^2]^{1/2}$, $b_{\pi/2} = Ze^2/(m_i v_s^2)$, $b_c = a(1+2b_{\pi/2}/a)^{1/2}$, and $\Gamma$ the standard Coulomb logarithm

$$\Gamma = \frac{1}{2}\ln\frac{\lambda_D^2 + b_{\pi/2}^2}{b_c^2 + b_{\pi/2}^2} \qquad (5)$$

For $F_i(2)$ the ion Debye length $\lambda_{Di}$ was used in (5), whereas for $F_i(3)$ the electron Debye length $\lambda_{De} = (T_e/T_i)^{1/2} \lambda_{Di}$. The electron temperature $T_e$, following from Langmuir probe measurements, is between 7.2 eV at 60 Pa and 7.7 eV at 20 Pa, leading to an electron Debye length between 1.3 mm and 1.7 mm. Using the electron Debye length instead of the ion Debye length leads to an enhancement of the ion drag force given by Barnes' formula of about a factor of 5 for the small particles up to a factor of 50 for the large ones.

| p[Pa] | $F_i$ (exper) | $F_i(1)$ | $F_i(2)$ | $F_i(3)$ |
|---|---|---|---|---|
| 20 | 6.4 | 4.4 | 2.6 | 11.0 |
| 40 | 4.4 | 1.9 | 1.5 | 7.3 |
| 60 | 4.1 | 1.3 | 1.1 | 5.4 |

Table 4: Comparison of the measured ion drag force (in $10^{-14}$ N) for particles with 1.28 µm diameter with theoretical predictions: (1) Ref.[9], (2) Ref.[8] with ion Debye length, (3) Ref.[8] with electron Debye length.

| p[Pa] | $F_i$ (exper) | $F_i(1)$ | $F_i(2)$ | $F_i(3)$ |
|---|---|---|---|---|
| 20 | 10.1 | 7.7 | 3.8 | 23 |
| 40 | 7.7 | 3.5 | 2.2 | 16 |
| 60 | 6.5 | 2.3 | 1.5 | 12 |

Table 5: Comparison of the measured ion drag force (in $10^{-14}$ N) for particles with 1.95 µm diameter with theoretical predictions: (1) Ref.[9], (2) Ref.[8] with ion Debye length, (3) Ref.[8] with electron Debye length.

| p[Pa] | $F_i$ (exper) | $F_i(1)$ | $F_i(2)$ | $F_i(3)$ |
|---|---|---|---|---|
| 20 | 11.3 | 10.8 | 4.6 | 36 |
| 40 | 9.1 | 5.1 | 2.6 | 26 |
| 60 | 7.4 | 3.2 | 1.7 | 18 |

Table 6: Comparison of the measured ion drag force (in $10^{-14}$ N) for particles with 2.55 μm diameter with theoretical predictions: (1) Ref.[9], (2) Ref.[8] with ion Debye length, (3) Ref.[8] with electron Debye length.

| p[Pa] | $F_i$ (exper) | $F_i(1)$ | $F_i(2)$ | $F_i(3)$ |
|---|---|---|---|---|
| 20 | 15.μ6 | 15.8 | 5.3 | 59 |
| 40 | 10.3 | 7.4 | 2.8 | 44 |
| 60 | 7.3 | 4.7 | 1.9 | 30 |

Table 7: Comparison of the measured ion drag force (in $10^{-14}$ N) for particles with 3.42 μm diameter with theoretical predictions: (1) Ref.[9], (2) Ref.[8] with ion Debye length, (3) Ref.[8] with electron Debye length.

| p[Pa] | $F_i$ (exper) | $F_i(1)$ | $F_i(2)$ | $F_i(3)$ |
|---|---|---|---|---|
| 20 | 25 | 31 | 5.9 | 184 |
| 40 | 21 | 16 | 2.8 | 135 |
| 60 | 16 | 9.1 | 1.9 | 90 |

Table 8: Comparison of the measured ion drag force (in $10^{-14}$ N) for particles with 6.86 μm diameter with theoretical predictions: (1) Ref.[9], (2) Ref.[8] with ion Debye length, (3) Ref.[8] with electron Debye length.

From these comparisons we arrive at the following conclusions:

(1) The best agreement between the ion drag measurements with PK-4 and theory is given by Khrapak's formula (2). This holds in particular at lower pressures, which is not surprising as ion collisions in the plasma are neglected in Khrapak's formula and ion-neutral collisions become important at higher pressures. The mean free path of the ions is comparable to the Debye length at pressures above 40 Pa.

(2) Barnes' formula (4) is excluded by the experiment. Using the ion Debye length, Barnes' formula underestimates the data clearly, while using the electron Debye length it overestimates the experiment significantly. The latter approach, however, is not justified anyway because the experiments took place in the bulk plasma with a subthermal drift velocity ($u = 130 - 390$ m/s, $v_{Ti} = 351$ m/s). Hence the Debye length is given by the ion Debye length [12].

(3) The size dependence of the ion drag force is in good agreement with theory. The experimental ion drag force decreases slower with increasing pressure than predicted, indicating the importance of ion-neutral collisions which lead to an enhancement of the ion drag force at higher pressures [16].

**Acknowledgments**: This work was supported by DLR under grant 50 WP 0204 and by ESA at the 36[th] ESA parabolic flight campaign.